\documentclass[twocolumn,3p,preprint]{elsarticle}
\usepackage[utf8]{inputenc}
\usepackage{hyperref}
\usepackage{graphicx}
\usepackage{lineno}
\biboptions{sort&compress}
\journal{Current Opinion in Systems Biology}

\begin{document}
\begin{frontmatter}

\title{Dynamical models for metabolomics data integration}

\author[1,2]{Polina Lakrisenko}
\author[1]{Daniel Weindl\corref{cor1}}
\ead{daniel.weindl@helmholtz-muenchen.de}
\cortext[cor1]{Corresponding author}
\address[1]{Institute of Computational Biology, Helmholtz Zentrum München, Neuherberg 85764, Germany}
\address[2]{Center for Mathematics, Technische Universität München, Garching 85748, Germany}

\begin{abstract}
As metabolomics datasets are becoming larger and more complex, there is an increasing need for model-based data integration and analysis to optimally leverage these data.
Dynamical models of metabolism allow for the integration of heterogeneous data and the analysis of dynamical phenotypes. 
Here, we review recent efforts in using dynamical metabolic models for data integration, focusing on approaches that are not restricted to steady-state measurements or that require flux distributions as inputs. Furthermore, we discuss recent advances and current challenges. We conclude that much progress has been made in various areas, such as the development of scalable simulation tools, and that, although challenges remain, dynamical modeling is a powerful tool for metabolomics data analysis that is not yet living up to its full potential.
\end{abstract}


\end{frontmatter}


\section*{Introduction}

Metabolism is a key determinant of cellular behavior, and metabolomics approaches are being applied in a wide range of domains \cite{JangChe2018}. Nowadays, metabolomics experiments yield rich datasets, which poses new challenges for their analysis. Current metabolomics assays provide information on hundreds to thousands of metabolites and large numbers of samples can be measured within reasonable times \cite{LiuZho2019}. Since holistic interpretation of the resulting datasets using 
``mental models'' becomes impossible, computational models are required and increasingly used for data integration and interpretation. However, observations from individual experiments are still commonly analyzed independently and in a qualitative manner, instead of being integrated in an overarching formal quantitative model.

A variety of metabolic modeling  frameworks have emerged, which allow for the integration of varying types of data, each with their specific advantages and disadvantages. A broad overview of these existing metabolic modeling approaches is provided in recent reviews \cite{VolkovaMat2020, Antoniewicz2021,FosterWan2021}.
In this article, we focus on metabolomics data integration using dynamical models. These models, usually specified in the form of ordinary differential equations (ODEs), are particularly appealing because: 1) they allow for both the integration of various types of metabolomics data and other types of data; 2) they allow for the analysis of inherently dynamical phenotypes; and 3) they can provide quantitative dynamical information on latent quantities, such as metabolic fluxes or compartment-resolved metabolite levels. On the downside, dynamical models come with a larger number of \textit{a priori} unknown parameters and are computationally more challenging, and therefore, are less scalable than other modeling approaches. When only considering steady-state measurements, these issues can be circumvented to a large extent, as exemplified by the recently developed K-FIT algorithm \cite{GopalakrishnanDas2020}. However, one is often interested in transient behavior, or a case where a steady-state cannot be attained. Therefore, although such methods have provided great insights in many applications, they are not discussed further here. We also exclude methods that require a known flux distribution or a biochemical objective function as input. For an overview of these methods and their applications, see, for example, \cite{FosterWan2021,StrutzMar2019,SaaNie2017}.

In the following, we give a brief overview of different types of metabolomics data, review recent examples of metabolomics data integration and analysis using dynamical models, and discuss major challenges and recent advances.

\section*{Metabolomics data}

Metabolomics measurements are usually performed using mass spectrometry, nuclear magnetic resonance, or enzymatic assays, with mass spectrometry  being the most widespread. Metabolomics datasets can be very rich and diverse, depending on the analytical platforms and experimental protocols. There are the classical label-free metabolomics approaches, which assess metabolite levels, and there are stable-isotope-assisted approaches, which determine isotopic enrichment after application of stable-isotope-labeled tracers \cite{JangChe2018}. 
Recent mass spectrometry methods can quantify isotopic enrichment of more than 100 metabolites in a single run \cite{ShiXi2020}. 
Various types of samples are analyzed, such as complete cell populations and extracellular media, and in some cases compartment-specific metabolite pools are accessible \cite{NonnenmacherPal2017}. Single-cell metabolomics assays have also emerged \cite{AliAbo2019,ThieleWun2019}. Measurements can be time-resolved or for single timepoints, quantitative or semi-quantitative. 
Due to this heterogeneity, various metabolic modeling approaches have been developed that are, more or less, tailored or restricted to specific types of data \cite{VolkovaMat2020}. However, in many studies, a combination of different types of data is acquired and integration of all these data is possible with dynamical models \cite{MillardEnj2021,RamosRat2020}.

\section*{Applications}

\begin{figure*}[ht!]
    \centering
    \includegraphics[scale=0.35]{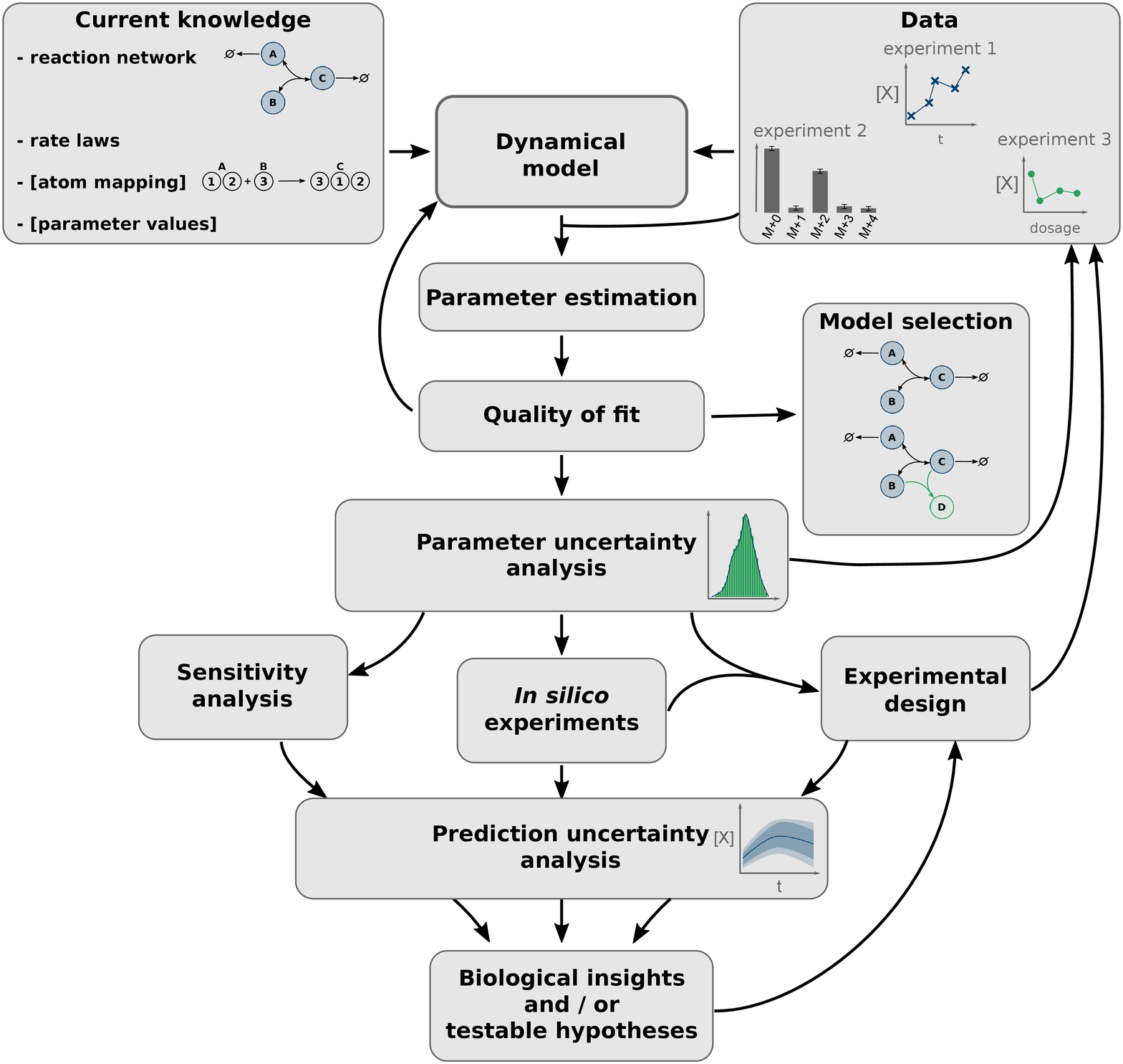}
    \caption{Typical steps for data integration using dynamical models. A model is developed based on current knowledge (brackets indicate optional components). Experimental observations of various types are used as model inputs or to estimate model parameters. If parameters are estimated, the model fit and parameter intervals are assessed.  Competing hypotheses or structural uncertainty can be encoded in different candidate models to select the most plausible one given the data. The parameterized model is used in various downstream analyses to derive biological insights. Examples include sensitivity analysis or metabolic control analysis to determine which parameters have the highest impact on reaction fluxes or metabolite concentrations. \textit{In silico} experiments can be performed to derive testable hypotheses and to design validation experiments. Considering uncertainties is crucial in all analyses to derive meaningful conclusions.}
    \label{fig:overview}
\end{figure*}

Model-based data analysis aims at analyzing experimental observations in the light of the current understanding of the observed system, as encoded in the model. 
Experimental observations are either used as model inputs or to infer model parameters. The behavior of the parameterized model is subsequently analyzed to obtain biological insights. Here, we focus mainly on the case of inferring model parameters from experimental data. Typical steps comprise 1) data acquisition, 2) model construction, 3) parameter inference, 4) uncertainty analysis, and 5) analyzing model fit and model-derived predictions (Fig.~\ref{fig:overview}).

Dynamical metabolic models have been applied in various contexts. Recurrently pursued objectives are 1) understanding dynamical processes \cite{LoevforsEks2020,BerndtBul2018}, for example, through comparison of competing models \cite{CarterFer2019}; 2) inferring control mechanisms and rate-limiting steps \cite{FeldmanSalitVei2019,MillardEnj2021}; and 3) leveraging those to push some system of interest in specific directions, for example, for strain optimization or drug target identification) \cite{MarinHernandezGal2020, OuBao2020, RamosRat2020}.

\citet{BerndtBul2018} developed a comparably large dynamical model of liver metabolism to, among others, study the response to various perturbations. The model describes the regulation of enzyme activities by allosteric effectors, hormone-dependent reversible phosphorylation, and variable protein abundances. A subset of parameters was estimated from measurements of intracellular and extracellular metabolite levels under 25 different experimental conditions. \citet{CarterFer2019} performed model selection among four candidate models, calibrated on  time-resolved measurements of metabolite levels,  to infer the most likely inhibitory mechanism in a drug-drug interaction. 
\citet{FeldmanSalitVei2019} trained a dynamical model of sulfur assimilation in \textit{Arabidopsis thaliana} on steady-state metabolite measurements, which allowed them to infer dynamical control patterns.
\citet{MillardEnj2021} analyzed time-resolved measurements of biomass, metabolite levels, and isotopic enrichment using a coarse-grained dynamical model that links glucose uptake to acetate metabolism and growth. Their analysis improved the understanding of control and regulation of acetate overflow and suggested the existence of a yet unknown regulatory program. 
\citet{MarinHernandezGal2020} applied a dynamical model of glycolysis, the pentose phosphate pathway, and glycogen metabolism of cancer and non-cancer cells, with parameters inferred from various enzyme assays, to derive suggestions for new therapeutic targets. \citet{OuBao2020} applied a dynamical model, which integrated experimental proteomics, metabolomics, and fermentation kinetics data, to identify key regulators and guide metabolic engineering of \textit{n}-butanol biosynthesis. Their comparison of the results from the dynamic model to those derived from a static model indicated higher accuracy of the former.
\citet{RamosRat2020} integrated both extracellular and intracellular time-resolved metabolite measurements from different experimental settings using a dynamical model comprising 33 state variables and describing central carbon metabolism and cell growth. They identified different physiological states and performed \textit{in silico} experiments to demonstrate how use of their model can improve cell line engineering and medium design.
\citet{MoonDon2020} used a dynamical model to predict the dynamics of mitochondrial NADPH concentration and NADPH/NADP$^+$ ratio in response to oxidative stress. Parameters of the model were estimated from time-resolved NADPH measurements. 
\citet{LoevforsEks2020} developed a model describing hormonal regulation of lipolysis. The model, comprising 15 state variables, was calibrated to time-resolved \textit{in vivo} and \textit{in vitro} measurements of metabolite and phosphoprotein abundance after different stimulations.
\citet{HorvathVil2020} calibrated an exceptionally comprehensive model of \textit{E. coli} cell-free protein synthesis, comprising 148 metabolites, based time-resolved on metabolite and protein product measurements. An ensemble model was subsequently used to identify the pathways with high influence on product yield.
\citet{YilmazPar2020} developed a coarse-grained model of protein synthesis in Chinese hamster ovary cells based on elementary flux modes. The model was calibrated to time-resolved measurements of cell density and levels of extracellular metabolites and protein product. 

These application examples demonstrate the potential of dynamical models to answer diverse research questions, by integrating and jointly analyzing various types of data. However, many current applications of dynamical models that integrate time-resolved measurements only cover small parts of the metabolic network or use very coarse-grained representations. This may be due to various challenges discussed in the following sections.

\section*{Constructing kinetic models}

Depending on the available data and research question, the model scope has to be defined and a model needs to be constructed. 
Model construction includes choosing and specifying 1) the network topology, and 2) kinetic rate laws. Retrieving or inferring model parameter values is covered in the following section.

Genome-scale metabolic reconstructions are a valuable resource for deriving the model topology. Such knowledgebases that describe the metabolic capabilities of an organism have been created for many species through automated reconstruction or extensive manual curation through community efforts \cite{FangLlo2020}. For example, the recently published Human1 \cite{RobinsonKoc2020} unifies two lineages of human genome-scale metabolic reconstructions and provides a good example of open and transparent curation. Although such reconstructions are continuously improved, structural uncertainty remains, for example due to enzymes of unknown function, enzyme promiscuity, or unclear  subcellular localization of enzymes.
Automated model reduction algorithms for generating targeted kinetic models from genome-scale metabolic reconstructions have been devised that can significantly speed-up model development \cite{vanRosmalenSmi2021,MasidAta2020}. However, care has to be taken, that the (over-)reduced model complexity does not skew analysis results \citet{HameriFen2021}.

If stable-isotope-labeling data are to be analyzed, the model not only needs to account for the pools and reactions of unlabeled metabolites, but also for those of the additionally occurring isotopic isomers (isotopologues). Deriving such isotopologue reaction networks requires knowledge of the mapping of substrate atoms to product atoms, for all reactions involving potentially-labeled compounds. A number of algorithms for automatically deriving such atom mappings have been developed (for example, \cite{JaworskiSzy2019}) and recent genome-scale metabolic reconstructions also include atom mappings \cite{BrunkSah2018}. However, correct atom mappings are crucial for modeling, and manual curation is still required, which is tedious for bigger models. The resulting isotopologue reaction networks can, depending on the network and tracer, be vastly larger than the usual metabolic networks due to the combinatorial complexity of stable isotope incorporation.

Having established a reaction network, mathematical expressions for describing reaction rates need to be chosen. Various approaches for specifying rate laws exist (reviewed in \cite{SaaNie2017}). The choice of rate laws is a trade-off between ``mechanisticness'' on the one hand and model complexity and number of parameters on the other hand.
Using a thermodynamics-based formalism \cite{GawthropPan2021,MasonCov2019} helps to create physically feasible models. Furthermore, such a parameterization can reduce the number of unknown parameters and simplify parameter estimation. The required thermodynamic parameters can be derived from experiments or approximated, by group contribution methods or quantum-mechanical simulations \cite{JoshiMcN2021}.

\section*{Determining model parameters}

Dynamical models usually come with a comparably high number of parameters whose values are not known \textit{a priori}. Some parameter values can be retrieved from databases, while others need to be estimated. Often, a combination of both approaches is used.

Commonly used databases of experimentally determined kinetic parameters include BRENDA \cite{ChangJes2020} and SABIO-RK \cite{WittigRey2017}. Additionally, the BioModels Parameters database \cite{GlontAra2020} provides structured access to parameter values that are used in models contained in the BioModels repository. If parameter values for a specific organism or experimental conditions are not available, values from a closely related species or settings may still be useful \cite{BerndtBul2018}. Datanator provides a simple interface to find such related parameter values based on various similarity measures \cite{RothLia2020}. However, using parameters determined by \textit{in vitro} assays or other makeshift parameter values comes with the caveat that they may not optimally reflect the \textit{in vivo} situation \cite{FosterWan2021}. As an alternative to taking these the parameter values from databases as true values, they can instead be used as prior information during parameter estimation \cite{SaaNie2017,FeldmanSalitVei2019}. 

Given sufficiently informative data, model parameters can be inferred, for example, via optimization- \cite{BerndtBul2018,CarterFer2019,FeldmanSalitVei2019,MillardEnj2021,MarinHernandezGal2020,OuBao2020,RamosRat2020,MoonDon2020,LoevforsEks2020,YilmazPar2020} or sampling-based \cite{HorvathVil2020} approaches. However, this is computationally costly as, for most cases, it involves thousands to millions of numerical ODE simulations.
Nevertheless, it has been demonstrated that parameter estimation is computationally tractable for large dynamical models. For example, through leveraging scalable algorithms, optimization-based maximum likelihood estimation was shown to be possible for a dynamical model of signaling pathways comprising over 1000 states and 4000 unknown parameters \cite{FroehlichKes2018}. In some cases, the structure of the optimization problem can be exploited to further improve convergence and reduce computational costs. This has been demonstrated for  optimization-based parameter inference from relative measurements \cite{SchmiesterSch2019a}, which are quite common in metabolomics datasets.

Scalable and highly optimized algorithms for model simulation and sensitivity analysis have been made available through easy-to-use toolboxes \cite{FroehlichWei2021, SomogyiBou2015} and there exists a wide variety of algorithms and tools for parameter inference \cite{MitraHla2019,VillaverdeFro2018}. Many of these tools are able to exploit increasingly available high-performance computing resources, which is key for moving towards larger dynamical models \cite{ChristodoulouLin2018,SchmiesterSch2019a,FroehlichKes2018}. The use of community standards for specifying models and parameter estimation problems \cite{SchreiberSom2020,SchmiesterSch2021,PorubskyGol2020} gives easy access to a majority of these tools. 

Sometimes model parameters are estimated independently and only later included in the full model \cite{MarinHernandezGal2020}. Although this can be computationally cheaper, it might not yield the best fit to the data.

\section*{Dealing with uncertainty}

Independently of the algorithm employed for parameter estimation, parameter estimates will be subject to uncertainty. This parameter uncertainty propagates to prediction uncertainty and can result in false conclusions \cite{VillaverdeRai2019}. Uncertainty analysis is, therefore, crucial.

Different methods to determine parameter confidence intervals exist. Commonly applied methods include those based on the Fisher information matrix (FIM), the profile likelihood approach, or sampling-based procedures \cite{WielandHau2021}. The choice of method is a trade-off between accuracy and computational complexity, with FIM-based methods being the cheapest computationally, but least accurate, and sampling being the most expensive but also most informative. Profile likelihood is often the method of choice, since it provides accurate results and is more scalable than sampling-based methods. Furthermore, as opposed to other methods, profile likelihood can be applied in spite of, and to detect, non-identifiable parameters \cite{WielandHau2021}.
Implementations of these algorithms are available through several easy-to-use software tools \cite{MitraHla2019}.

Methods that are similar to those used for parameter uncertainty analysis are applied for prediction uncertainty analysis \cite{VillaverdeRai2019,WielandHau2021}. Sampling-based approaches account for the prediction uncertainty by design, as a sample from the prediction posterior is acquired while generating a sample from the parameter posterior.
Where sampling or prediction profile likelihood are computationally too expensive, a cheaper ensemble-based approach can be applied instead \cite{VillaverdeRai2019}. During parameter estimation, multiple parameter vectors will be obtained that result in similarly good model fits. An ensemble of models can, for example, be built from these parameter vectors and the spread of the resulting simulations can be used as an estimate for prediction uncertainty \cite{VillaverdeRai2019}.

Recent applications of dynamical models of metabolism include examples of identifiability analysis \cite{RoyFin2019,CarterFer2019}, sampling-based assessment of parameter and prediction uncertainties \cite{MillardEnj2021}, as well as other approaches uncertainty analyses \cite{LoevforsEks2020}.

It is important to note that large parameter uncertainties do not have to manifest in large prediction uncertainties, but may still allow for deriving robust predictions  \cite{FeldmanSalitVei2019,FroehlichKes2018,ChristodoulouLin2018}. 
For example, \citet{FeldmanSalitVei2019} generated an ensemble of models that reproduce the available data equally well and used it to derive predictions that were consistent across the ensemble.

\section*{Conclusions}

Increasingly detailed metabolic reconstructions for more and more organisms provide the basis for constructing genome-scale models, but also for deriving targeted sub-models.
Efficient implementations of scalable algorithms enable simulation and sensitivity analysis for dynamical models with as many as a few thousand state variables, even on personal computers.
Adopting community standard for specifying models or parameter estimation problems not only facilitates reproducibility and reusability of models and data \cite{PorubskyGol2020}, but also grants easy access to a wide ecosystem of tools for model simulation, parameter inference and further analyses.

Computationally challenging parameter estimation is often considered the major bottleneck for data-based dynamical (metabolic) modeling \cite{FosterWan2021}. It is true that for very large models this still is intractable and more scalable algorithms are required. Results from other fields of application suggest that a better leveraging of both existing tools and computational resources can facilitate work with models and datasets that are an order of magnitude larger than those currently used in the field of dynamical modeling of metabolism.

Aside from any algorithms, successful model-based data analysis depends on matching models and data. To obtain informative data, experiments should ideally be designed with a specific modeling goal in mind, and the available data should be taken into account when building models. Nevertheless, parameter uncertainty will always remain.  
Robust predictions may still be possible despite large parameter uncertainties. Therefore, dynamical modeling should not be ruled out prematurely due to supposedly too sparse data. However, awareness and assessment of parameter and prediction uncertainties is crucial, although often neglected. 

In summary, we feel that dynamical modeling is a valuable, yet underused, tool for integrative metabolomics data analysis that could help to derive a more comprehensive, quantitative, and dynamical understanding of metabolism in many applications.

\section*{Acknowledgements}

This work was supported by the German Federal Ministry of Education and Research (BMBF) within the e:Med funding scheme (junior research alliance PeriNAA, grant no. 01ZX1916A).

\bibliographystyle{elsarticle-num-names}
\bibliography{review.bib}
\end{document}